\documentclass[aps, pra, twocolumn, superscriptaddress, 10pt, floatfix]%
              {revtex4-1}
\usepackage{calc}
\usepackage{xcolor}
\usepackage{braket}
\usepackage{amsmath}
\usepackage{siunitx}
\usepackage{graphicx}
\usepackage{mathtools}
\usepackage{graphicx}
\usepackage{hyperref}

\begin{document}

\title{Timing quantum emission: Coherence, superradiance, and entanglement in order}

\author{Nur Fadhillah Binti Rahimi}       \email{rahnu978@student.otago.ac.nz}
\affiliation{Dodd-Walls Centre for Photonic and Quantum Technologies, Department of Physics, University of Otago, Dunedin 9016, New Zealand}

\author{Norman Tze Wei Koo}         
\affiliation{Centre for Quantum Technologies, National University of Singapore, Singapore 117543, Singapore}

\author{Daniel Schumayer}           \email{daniel.schumayer@sydney.edu.au}
\affiliation{Dodd-Walls Centre for Photonic and Quantum Technologies, Department of Physics, University of Otago, Dunedin 9016, New Zealand}
\affiliation{School of Physics, University of Sydney, Camperdown NSW 2050, Australia}

\author{Christopher Gies}           
\affiliation{Institute for Physics, Faculty V, Carl von Ossietzky University Oldenburg, 26129 Oldenburg, Germany}

\author{Leong Chuan Kwek}           
\affiliation{Centre for Quantum Technologies, National University of Singapore, Singapore 117543, Singapore}
\affiliation{Institute of Advanced Studies, Nanyang Technological University, Singapore 639798, Singapore}
\affiliation{National Institute of Education, Nanyang Technological University, Singapore 637616, Singapore}
\affiliation{MajuLab, CNRS-UNS-NUS-NTU International Joint Research Unit, Singapore 117543, Singapore}

\author{David A.~W. Hutchinson}     
\affiliation{Dodd-Walls Centre for Photonic and Quantum Technologies, Department of Physics, University of Otago, Dunedin 9016, New Zealand}
\affiliation{Centre for Quantum Technologies, National University of Singapore, Singapore 117543, Singapore}

\begin{abstract}
    We investigate the short-term temporal dynamics of superradiance in closely spaced quantum emitters. Building on Dicke’s 1954 framework, we analyze the sequential emergence of coherence, superradiance, and entanglement, revealing a distinct temporal hierarchy in their extremal values: relative coherence develops first, followed by the peak of correlated emission, then minimal entanglement, and subsequently minimal spin-spin correlation is reached. These findings suggest that enhanced relative coherence precedes correlated emission, and when spin correlations are negligible, entanglement and correlated emission become linked in time.
\end{abstract}

\date{\today}
\maketitle

\section{Introduction}

In 1954 Dicke proposed a novel description of spontaneous light emission in a system of particles coupled to an electromagnetic field \cite{Dicke1954}. Previously, the concept of light emission had been described as radiation of photons emitted by independent particles, which is indeed justified if the typical distance between randomly distributed particles is larger than the wavelength of emitted light waves. However, if the close proximity of particles cannot be neglected, spontaneous light emission is better described as a collective, self-organization phenomenon \cite{Hepp1973}: particles become correlated due to each other's radiation field.

\begin{figure}[b!]
    \includegraphics[width=0.7\linewidth]{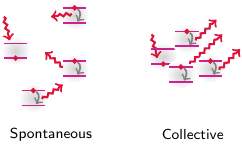}
    \caption{\label{Fig:Schematics}
             (Color online) Schematics of spontaneous radiation and superradiant emission. On the left, emitters are far apart and decay individually following a characteristic rate, while on the right the emitters are influenced by each other's radiation field, hence are constructively coupled. 
            }
\end{figure}

A system of two-level emitters, archetypically represented by spin-$\tfrac{1}{2}$ particles, that are prepared in their individual excited states eventually decays to the ground state by emitting photons spontaneously. For independent emitters the decay rate is proportional to the number of emitters, $N$ \cite{Dicke1954}, while for correlated emitters it scales as $N^2$, which is a well-established signature of superradiant emission \cite{mok_dicke_2023}. Consequently, the peak photon emission rate appears as a sudden burst of radiation, hence the term ``superradiance,'' which fundamentally relies on the correlations among emitters, as illustrated schematically in Fig.~\ref{Fig:Schematics}.

Superradiance and superfluorescence \footnote{Superradiance and superfluorescence are similar phenomena, but in the former case a macroscopic dipole moment is created at the initialization, while in the latter case -- due to incoherent initialization -- no such dipole moment is present by the end of the initialization. Hence in superradiance there is an experimentally controlled direction (i.e., given by the net dipole moment), which also influences the observed effect.}, can be highly beneficial for quantum-enhanced measurements and quantum metrology \cite{Paulisch2019}: a steady-state superradiant lasing with its ultra-narrow line-width \cite{Kristensen2023, Barberena2023} would improve optical clock frequencies \cite{Norcia2016} and spectroscopically relevant transitions \cite{Norcia2016, Norcia2018, Laske2019}, or allow for terahertz amplifiers \cite{Masahiro2003}, optical emitters \cite{Teperik2012}. Furthermore, the critical behavior near superradiant phase transitions permits reaching Heisenberg limit in measurements \cite{Wang2014}. 

Superradiance is not only useful because the peak intensity scales with $N^{2}$, but also due to its tunability. It was first demonstrated in thermal gases \cite{Skribanowitz1973, Gross1976}, and later in single diamond nanocrystals \cite{Bradac2017}, while superfluorescence was observed in spatially inhomogeneous systems, in colloidal perovskite nanocrystals \cite{Raino2018, Nguyen2023}. Recently superfluoresent emitters were coupled to nanoparticles at room temperature \cite{Huang2022}; an immense step away from cryogenic experimental setups towards applications. An intense laser pulse (\si{\nano\second} or \si{\femto\second}) excites many ions within a single nanoparticle forming a collective quantum state that emits light faster than usual: emission lifetime shortens from \si{\micro\second} to \si{\nano\second}. In semiconductor systems, a recombination-time shortening by a factor of 40, down to \SI{20}{\pico\second}, has been reported \cite{Jahnke2016}.

Nowadays, it is instructive to understand how and when a system reaches the superradiant state, and to quantify how superradiance, coherence, and entanglement relate during evolution. Earlier, we established \cite{Lohof2023} a close connection between emitter correlations and an entanglement witness sensitive to Dicke-states. From a quantum resource-theory perspective, it also seemed natural to ask: what drives superradiance? Does coherence build up and then get converted into superradiance, or is entanglement, as the resource, converted into coherence or superradiance? Such questions are far from trivial. Given a traditional Dicke setup with fully inverted emitters, two independent studies form a complementary understanding on superradiance and entanglement. While Rosario \emph{et al.} \cite{Rosario2025} deduced that entanglement is not a prerequisite for superradiance, Bassler \cite{Bassler2025} has proven that the traditional setup, while undergoing superradiance, does not generate entanglement. Ferioli \emph{et al.} \cite{Ferioli2025} explored that coherence varies during superradiant emission: both first and second-order coherence emerge during the initial development of superradiance peak, but subsequently drop and indicate anticorrelations (photons are unlikely to be emitted simultaneously) after which it settles to a stable value (subradiant phase). Furthermore, local observables can be well-approximated via a mean-field approach, even if a considerable amount of entanglement is generated in super- and subradiant decay processes \cite{Zhang2025}. In this paper, we investigate the temporal characteristics: in what sequence coherence, entanglement and superradiance build up.

\section{Formalism}

The original Dicke model qualitatively explains superradiance, but it is insufficient to describe realistic systems, as it considers only the decay of emitters. In practice, the emitters' dynamics and interaction with the electromagnetic field are intertwined. The model also assumes that emitters are much closer than the emitted photon’s wavelength which requires the inclusion of dipole-dipole interactions. If this interaction, combined with collective decay, is sufficiently strong, it may break the global symmetry (conservation of parity of the total number of excitations), and may lead to a phase transition from the normal to the superradiant state \cite{Lambert2004, Bastidas2012, Kirton2019, Roses2020, Das2024}.

The Tavis-Cummings Hamiltonian \cite{Garraway2011, Kirton2019} governs $N$ identical two-level emitters coupled to photons
\begin{equation} \label{eq:TavisCummingsHamiltonian}
    H
    =
    \tfrac{1}{2} \omega_{q} \sum_{i=1}^{N} \sigma^{z}_{i} +
    g \sum_{i=1}^{N} \left (\sigma^{+}_{i} \alpha + \sigma^{-}_{i} \alpha^{\dagger} \right) +
    \omega_c \alpha^{\dagger} \alpha,
\end{equation}
where $\omega_{q}$ is the resonance frequency of an emitter, $g$ is the coupling strength of an emitter and a photon, while $\omega_{c}$ is the angular frequency of a photon. The operators $\sigma^{z}_{i}$ and $\sigma^{\pm}_{i} = \tfrac{1}{2}(\sigma^{x}_{i} \pm i\sigma^{y}_{i})$ are the inversion, raising and lowering operators for emitter $i$, while $\alpha^{\dagger}$ and $\alpha$ are creating and annihilating photons in the cavity mode.

In an open quantum system, incoherent processes, such as decay, photon loss and pure dephasing at rates $\gamma$, $\kappa$ and $\gamma_{\phi}$, can be captured within a Markovian approximation and the Lindblad master equation
\begin{align} \label{eq:LindbladMasterEquation}
    \frac{d}{dt} \rho
    =
    - i \left [ H, \rho \right ] + 
    \sum_{k \in \lbrace \gamma, \kappa, \gamma_{\phi}\rbrace}{\!\!\!D_{k}(\rho)},
\end{align}
where $D_{k}$ follows the standard Lindbladian structure
\begin{align*}
    D_{k}^{\phantom{\dag}}(\rho)
    =
    k
    \left \lbrack
        L_{k}^{\phantom{\dagger}} \rho L_{k}^{\dagger} -
        \frac{1}{2} \left \lbrace
                        L_{k}^{\dagger} L_{k}^{\phantom{\dagger}}, \rho 
                    \right \rbrace
    \right \rbrack
\end{align*}
with $L_{\gamma} = \sigma^{-} \otimes \mathrm{I}$, $L_{\kappa} = \mathrm{I} \otimes \alpha$, and $L_{\gamma_{\phi}} = \sigma^{z} \otimes \mathrm{I}$. 

Here we solve Eq.~\eqref{eq:LindbladMasterEquation} using two approaches. Exact calculation for $N < 10$ is feasible and suits experimental setups based on superconducting qubits, quantum dots \cite{Lu2023a, Lu2023b}, or the silicon-on-insulator platform \cite{Liu2014}. If, however, emitters are indistinguishable and permutation invariance holds, the calculation extends to $N \le 60$ using an optimized numerical approach  \cite{Shammah2018}. We consider the situation in which all emitters occupy the deexcited ground states, while $\tfrac{1}{2}N$ photons populate the resonant mode of the quantized electromagnetic field \cite{johansson_qutip_2012, johansson_qutip_2013}.

We investigate the temporal evolution of correlated emission, $C_{0} = \braket{\sigma_{i}^{+} \sigma_{j}^{-}}$, the von Neumann entropy, $S(\rho) = -\text{Tr}(\rho \ln{\rho})$, the relative entropy of coherence \cite{Baumgratz2014}, $C_{\text{rel}} = S(\rho_{\text{diag}}) - S(\rho)$, and the entanglement witness, $\braket{W} = 1 - 4\text{Re}(C_{0}) + C_{zz}$ \cite{Krammer2009}. Note that $C_{0}$ is related to $W$ via spin-spin correlation function, $C_{zz} = \braket{\sigma_{i}^{z} \sigma_{j}^{z}}$ ($i \ne j$). We note that $C_{0}$ and $S$ are basis independent, while $\braket{W}$ and $C_{\text{rel}}$ are not.
\begin{figure}[h]
    \includegraphics[width=83mm]{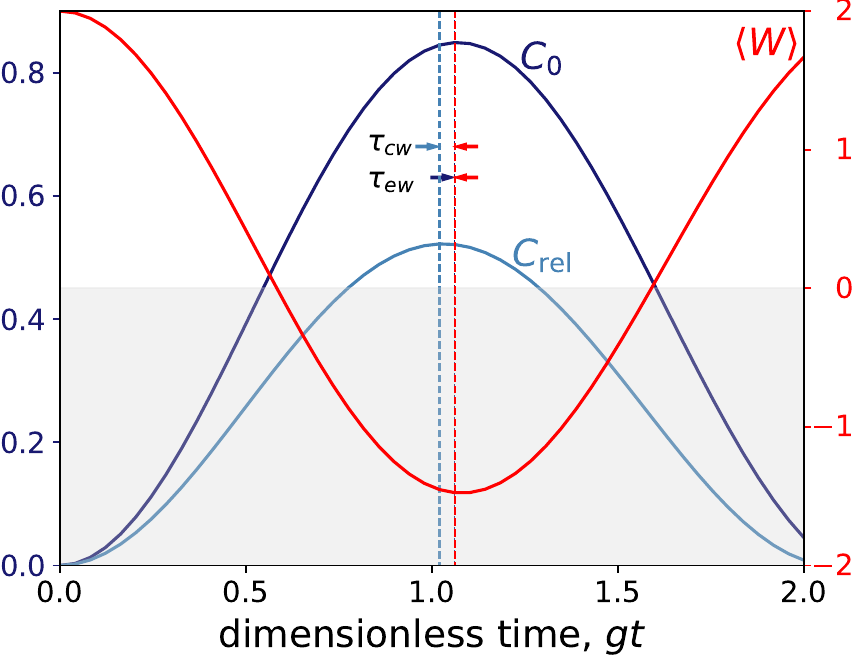}
    \caption{\label{fig:ExampleForTwoEmitters}
            (Color online) Correlated emission, relative entropy of coherence, and entanglement witness are shown for the $N=2$ emitter subsystem as functions of dimensionless time. Correlated emission and relative entropy of coherence are measured on the left ordinate, while entanglement witness is measured on the right ordinate (see colors). Vertical lines indicate those moments in time, where each quantity reaches its extremal value. In this setup $C_{\text{rel}}$, reaches its maximum first, then $C_{0}$ and $\braket{W}$ reach their extrema simultaneously. The corresponding time differences are $\tau_{\text{cw}}$ (coherence vs. witness) and $\tau_{\text{ew}}$ (emission vs. witness). Other parameters: $g=1$, $\gamma/g=0.1$, $\kappa/g=0.1$, and $\gamma_{\phi}/g=0.0225$.
           }
\end{figure}

\section{Timescales during superradiant emission}

There are earlier analyses of temporal evolution of superradiance; e.g., \citet{Haake1981} used passage time marking the moment when the emitted field first reaches a defined intensity, while \citet{Vrehen1981} investigated the delay time associated with the onset of superfluorescent pulses and its dependence on sample parameters, such as the Fresnel number, together providing complementary insights into the timing processes that govern the emergence of superradiant behavior. Here we define a characteristic time for each coherence measures as the first moment when they reach their extremal value after released from their initial state. Within the independent, single-particle picture the emitters decay at a linear rate, hence the characteristic timescale is $\tau \propto \tfrac{1}{N}$ \cite{andreev_collective_1980}. If, however, phase dispersion is approximately zero and particles exhibit collective behavior, the system reaches phase synchronization culminating in superradiant emission, which is associated with a characteristic time \cite{Nefedkin2017, Koppenhofer2022}
\begin{align*}
    \tau_{C_{0}} \approx \tau \ln{\!(N)} \approx \frac{\ln{\!(N)}}{N}.
\end{align*}
Since $\tau_{C_{0}} \propto N^{-1}$, the larger the system the sooner it becomes superradiant. Extremal phase dispersion and maximum radiation intensity occur simultaneously, reinforcing that phase synchronization is the mechanism that gives rise to superradiance. This behavior may be advantageous for quantum metrology purposes \cite{Koppenhofer2022, young_engineering_2024}. 
\begin{figure}[h]
    \includegraphics[width=83mm]{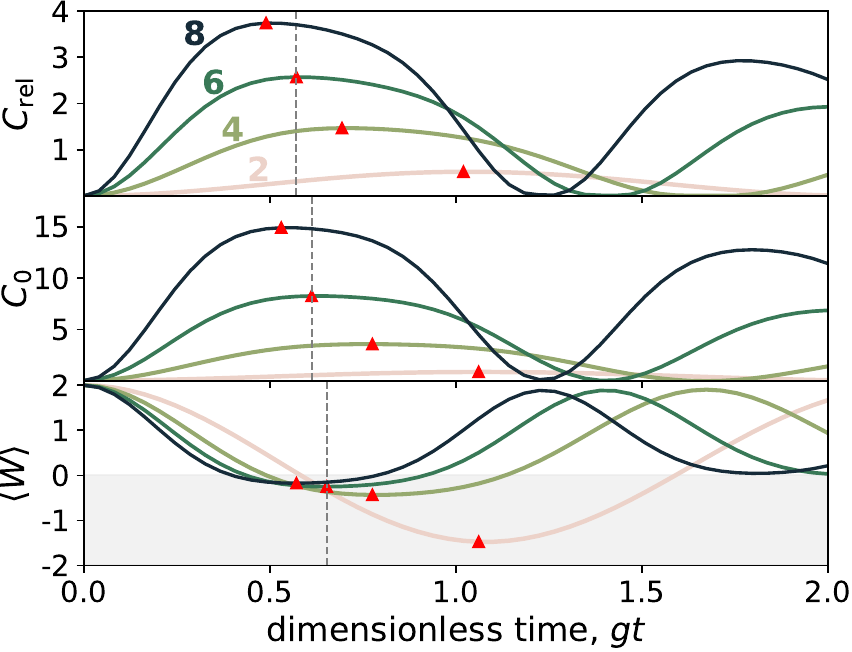}
    \caption{\label{Fig:SmallSystem_EvolutionAndExtrema}
             (Color online) Time evolution of $C_{\text{rel}}$, $C_{0}$, and $\braket{W}$, are plotted for $N=2$, 4, 6, and 8 emitters. The number of emitters are shown in the top figure as labels and different colors. The colors are common in all subplots.
             Vertical dashed lines indicate the moments when these quantities reach their extrema for $N=6$. Other parameters are as in Fig.~\ref{fig:ExampleForTwoEmitters}.
            }
\end{figure}

Figure~\ref{Fig:SmallSystem_EvolutionAndExtrema}(a) shows the evolution of relative coherence, correlated emission, and the expectation value of the entanglement witness for small systems ($N=2$, 4, 6, 8). All measures reach their extremal values, denoted by small red triangles, earlier as system size increases---a well-known result for $C_{0}$. Additionally, for fixed $N$ the coherence, $C_{\text{rel}}$ builds up first, followed by correlated emission, $C_{0}$, and finally entanglement. These moments when each of these quantities peak are marked by vertical dashed lines for $N=6$. Based on this observation we conjecture that
\begin{equation*}
    \tau_{C_{\text{rel}}} \le \tau_{C_{0}} \le \tau_{W}.
\end{equation*}
Equality can happen, e.g., for $N=2$, $\tau_{C_{\text{rel}}} < \tau_{C_{0}} = \tau_{W}$. Note that the inequality $\tau_{C_\text{rel}} < \tau_{W}$ could be indicative of a potential causal relationship between relative coherence and entanglement. This finding supports a prior mathematical study \cite{Streltsov2015}: coherence can be converted into entanglement. Yet, further microscopic analysis is needed to uncover the process of such conversion.
\begin{figure}[h]
    \includegraphics[width=84mm]{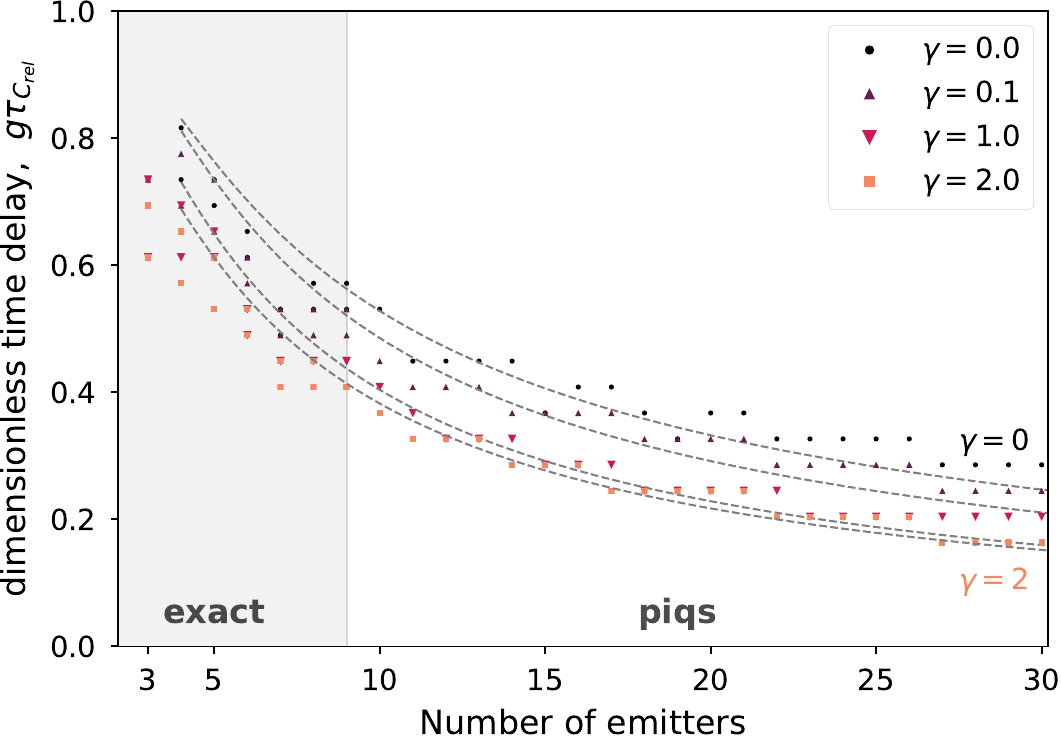}
    \caption{\label{Fig:TimeDelay}
             (Color online) Time-delay, $\tau_{C_{\text{rel}}}$ is depicted as a function of $N$. The shaded area, $N \le 9$, is calculated using two different approaches: exact time-evolution in computational basis and a permutational invariant quantum solver (PIQS). These approaches lead to similar values. The four sets of data, distinguished by different markers and colors, correspond to different decay parameters, $\gamma = 0$, 0.1, 1, 2. Qualitatively, $\tau_{C_{\text{rel}}}$ diminishes as $\ln{\!(N)}/N^{\alpha}$, where $\alpha \approx 1$ for $\gamma=0$, and $\alpha \approx 1.2$ for $\gamma = 2$, as shown by the dashed lines. Similar results apply to $\tau_{C_{0}}$ and $\tau_{W}$, but left out for clarity.
            }
\end{figure}
Fig.~\ref{Fig:TimeDelay} depicts the expected qualitative behavior of dimensionless time delay $g\tau_{C_{\text{rel}}}$ as a function of $N$ and $\gamma$; with increasing $N$ the time delay decreases as $\ln{(N)}/N^{\alpha}$, where $\alpha$ is a fitted parameter assumed to be weakly dependent on $\gamma$. Indeed for no emitter decay we recover the classical result $\alpha=1$. However, with increasing dissipative process, the reduction of time delay is slightly more pronounced, $\alpha\approx 1.2$ for strong decay. The case is qualitatively similar to that of a damped harmonic oscillator, whose first extremal deflection shifts to earlier times $\tau_{1} \propto (\omega_{0}^{2} - \beta^{2})^{-1/2}$ with increasing damping $\beta$. Here, $\omega_{0}$ is the angular frequency of the undamped harmonic oscillator. 

\begin{figure}[th!]
    \includegraphics[width=84mm]{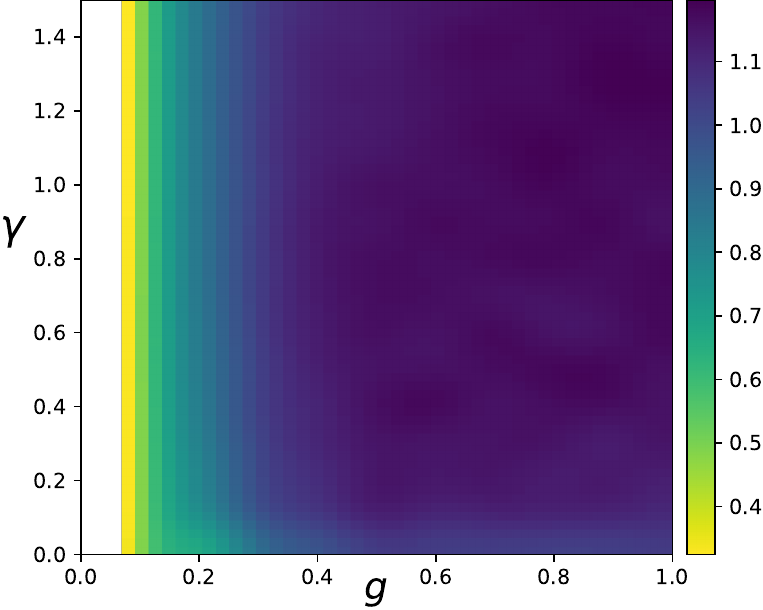}
    \caption{\label{Fig:Alpha}
             (Color online) Scaling exponent, $\alpha$, is depicted as a function of interaction strength $g$ and decay rate $\gamma$. White region corresponds to parameter combinations where the scaling exponent could not be reliably computed.
            }
\end{figure}

In order to capture its variability, $\alpha$ values are depicted as a heat map in Fig.~\ref{Fig:Alpha} in terms of $g$ and $\gamma$. We display in white the parameter combinations where the scaling exponent could not be determined, highlighting the regimes in which the scaling behavior is not well defined. One may deduce that for the current initial state, with $\tfrac{1}{2}N$ photons in the cavity, superradiance is generally observed at strong coupling ($\approx 0.5 \le g$), with the scaling exponent for the time delay ranging as $\alpha \in [1, 1.2]$.

\begin{figure}[hbt!]
    \includegraphics[width=83mm]{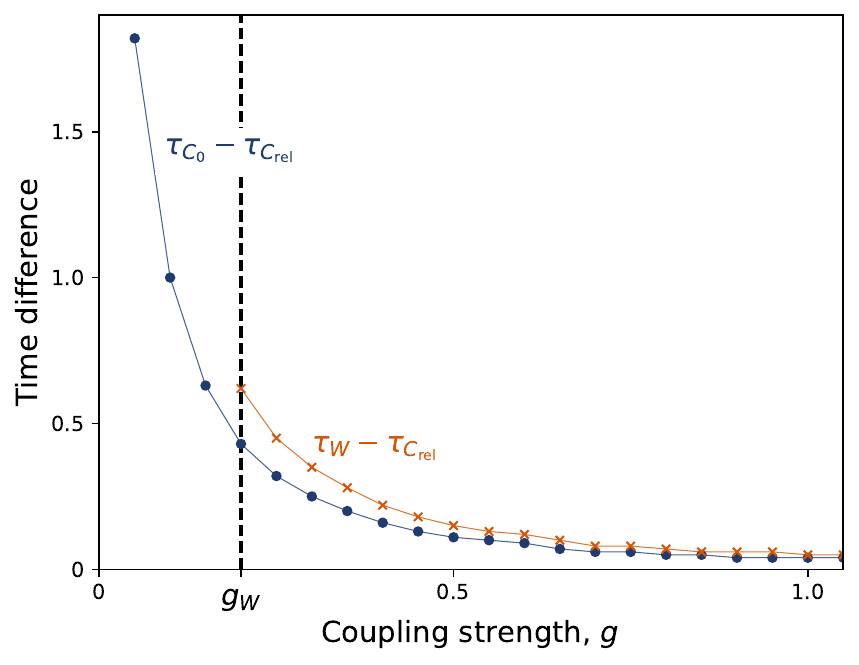}
    \caption{\label{fig:TimeGapsForTwoEmitters}
            (Color online) Time gaps, $\tau_{C_{0}} - \tau_{C_{\text{rel}}}$ and $\tau_{W} - \tau_{C_{\text{rel}}}$ are displayed for $N=2$ as functions of $g$. The vertical dashed line ($g_W \approx 0.20$) represents the interaction strength above which entanglement is detected. Other parameters: $n_{\text{p}} = 1$, $\gamma = 0.1$, $\kappa = 0.1$, and $\gamma_{\phi} = 0.0225$.
           }
\end{figure}

In Fig.~\ref{fig:TimeGapsForTwoEmitters} the behavior of  time differences between three measures are studied as a function of the light-matter coupling strength, keeping $C_{\text{rel}}$ as the baseline. We find that $\tau_{W}$ is always larger than $\tau_{C_{0}}$, and also that any kind of time difference vanishes as the coupling strength increases. The latter effect is expected; for increasing $g$ the system exhibits faster Rabi oscillations between collective atomic states and the cavity mode, and $\rho$ also evolves more rapidly, reflecting quicker population transfers. Its off-diagonal elements (coherences) decay faster if dephasing is present, hence all $C_{0}$, $C_{\text{rel}}$, and $C_{zz}$ diminish.

The order of $\tau_{C_{\text{rel}}}$ and $\tau_{C_{0}}$ lies in their operational definitions: $C_{\text{rel}}$ is an entropy-based measure which detects the purely incoherent part of the density operator, while $C_{0} = \braket{\sigma_{i}^{+} \sigma_{j}^{-}}$ can only be non-zero if the density operator has off-diagonal elements. Thus, enhanced relative coherence precede correlated emission for the evolution of all initial states with the exception of cases where the maxima of both quantities are not well defined (refer to Fig.~\ref{fig:appendix_separability}(c) of Appendixes). $\braket{W}$ is related to $C_{0}$ and $C_{zz}$; for $C_{zz} \approx 0$, the entanglement is in one-to-one correspondence with $C_{0}$ and their time gap vanishes. However, one may not conclude on the relative timings in the general case, when $C_{zz}$ is not negligible. Finally, Fig.~\ref{fig:TimeGapsForTwoEmitters} suggests that the time gaps between relative coherence and correlated emission, $\tau_{C_{0}} - \tau_{C_{\text{rel}}}$, and the time gaps between relative coherence and entanglement witness, $\tau_{W} - \tau_{C_{\text{rel}}}$, diminish as the interaction increases. 

\begin{figure}[bt]
    \includegraphics[width=84mm]{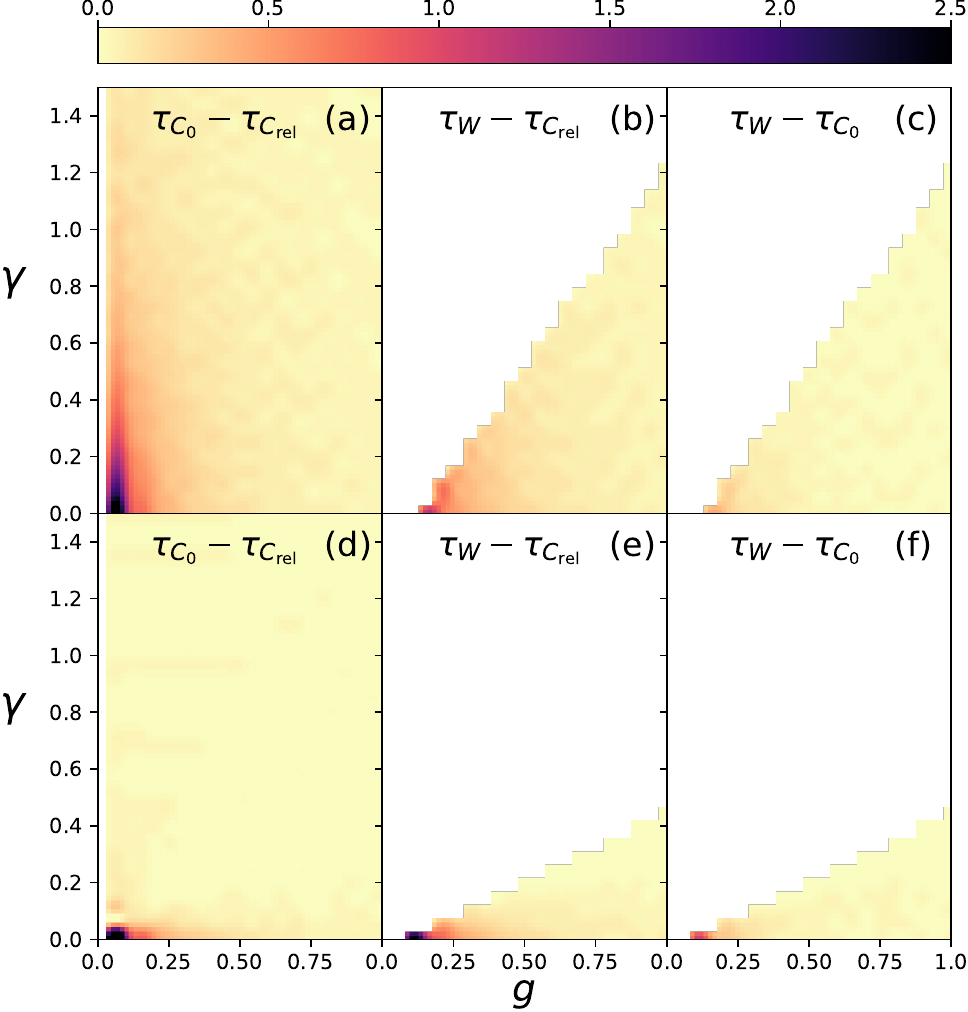}
    \caption{\label{fig:Fig7}
             Pairwise time gaps are shown: (a, d) $\tau_{C_0} - \tau_{C_{\text{rel}}}$, (b, e) $\tau_{W} - \tau_{C_{\text{rel}}}$, and (c, f) $\tau_{W} - \tau_{C_0}$. The difference between the two rows is the initial state. In top row $(N_{\text{ex}}, n_{\text{p}}) = (0, 1)$, while in the bottom row  $(N_{\text{ex}}, n_{\text{p}}) = (2, 0)$. Other parameter values are shared over all subplots: $\kappa =0.1$, $\gamma_{\phi} = 0.0225$, $\omega_{q} = 100$, $\omega_{c} = 100$, and $t_\text{max} = 20.0$. Large white regions in (b), (c), (e) and (f) indicate no entanglement witnessed.
            }
\end{figure}

To clarify this observation, we plotted the time gaps as functions of the interaction strength $g$ and decay rate $\gamma$ for distinct initial conditions. In the captions, $N_{\text{ex}}$ denotes the number of emitters prepared in their excited states, and $n_{\text{p}}$ is the number of photons present initially in the cavity. Note that the number of emitters in ground states is $N - N_{\text{ex}}$. Top row of Fig.~\ref{fig:Fig7} corresponds to $(N_{\text{ex}}, n_{p}) = (0, 1)$, whereas in the bottom row the system was prepared in the fully inverted Dicke state, $(N_{\text{ex}}, n_{p}) = (2, 0)$. First, one may conclude that the time gaps seem to be always positive irrespectively of the initial state and of the two dynamic parameters, $g$ or $\gamma$. Another generic feature is that all time gaps are pronounced in the weak coupling and weak decay limit. Furthermore, for any coupling strength, there exists a decay value above which the entanglement witness is never negative, as indicated by the nearly diagonal line in Fig.~\ref{fig:Fig7}(b, c, e, and f). Interestingly, preparing the system in the $(N_{\text{ex}}, n_{p}) = (0,1)$ initial state exhibits a bigger region in the ($g$, $\gamma$) space, where the $\tau_{C_{0}} - \tau_{C_{\text{rel}}}$ time gap is mostly controlled by $\gamma$. This feature is qualitatively understandable: initializing the emitters in their ground state, it takes time for the photons to excite these emitters and then reemit the photons in a superradiant manner. Small decay rate means that this reemission takes long time, meanwhile relative coherence can be built up.

\section{Conclusion}

We investigated the time evolution of superradiance, coherence, entanglement, and the interplay of their associated time scales, showing that for uncorrelated initial states, these quantities have a clear temporal order: coherence builds up first, followed by the correlated emission peak, and then the witness becomes negative, indicating entanglement. In summary:
\begin{equation} \label{eq:InequalityInConclusion}
    \tau_{C_{\text{rel}}} \le \tau_{C_{0}} \le \tau_{W} \le \tau_{C_{zz}} .
\end{equation}
We have extended our analysis to a variety of initial states, the results of which are shown in the Appendixes. The inequalities in \eqref{eq:InequalityInConclusion} seems to be robust and stand for a wide class of initial states.

In light of a number of recent investigations of the sub- and superradiance decay dynamics \cite{Lohof2023, Bassler2025, Rosario2025, Ferioli2025, Zhang2025}, some of which are demonstrating that the current experimental capability is already sufficient to test our prediction, we expect that our results can be readily and directly verified in small systems in the near future.

Beyond these results, our analysis also highlights a conceptual connection that goes beyond mere structural features of the chosen metrics. The quantity $C_{0}$ is rooted directly in the light–matter interaction responsible for collective emission, and its behavior reflects a genuine physical mechanism. By contrast, $C_{\text{rel}}$ is a resource‑theoretic measure whose meaning becomes physically grounded only once the basis is fixed. Here the energy eigenbasis lends itself naturally by the Hamiltonian. The fact that these two fundamentally different quantities exhibit a consistent temporal relationship, at least within a class of initial conditions, suggests that the observed hierarchy is not accidental but reflects an underlying dynamical structure of the emission process. Finally, while our entanglement analysis relies on a geometric witness rather than a full resource‑theoretic measure, this points toward a promising direction for future work, where a resource-theoretic entanglement measure may reveal even deeper connections.

\acknowledgments
Support from the University of Otago, the Dodd-Walls Centre and Quantum Technologies Aotearoa is appreciatively recognized by N. F.B.R. C. G. gratefully acknowledges financial support from the German Federal Ministry of Research, Technology and Space (BMFTR) via the projects QR.X and QR.N (16KISQ024,16KIS2203). 

\section*{Data Availability} The datasets corresponding to Figs.~2--9 of this paper are available in the Zenodo repository \cite{Binti2026Data}. The software used to generate the data are all open source and cited within the paper.

\appendix

\section{Effect of initial state}

In the main text, we analyzed a composite system consisting of spin-$\tfrac{1}{2}$ particles prepared in their ground state and a single‑mode cavity initially containing $n_{p}$ photons. As this initialization differs from the conventional Dicke model, it is natural to ask whether the resulting hierarchy of characteristic delays persists under alternative initial conditions. We examined a few additional initial configurations in which some emitters are prepared in their excited states, while the cavity may also contain some photons. The simulations confirm that the characteristic temporal ordering and collective emission dynamics identified in the main article persist in most, but not all, cases. Thereby, the time-order in Eq.~\eqref{eq:InequalityInConclusion} seems to be a robust feature.

\begin{figure*}[t] \centering 
    \includegraphics[width=0.4\textwidth]{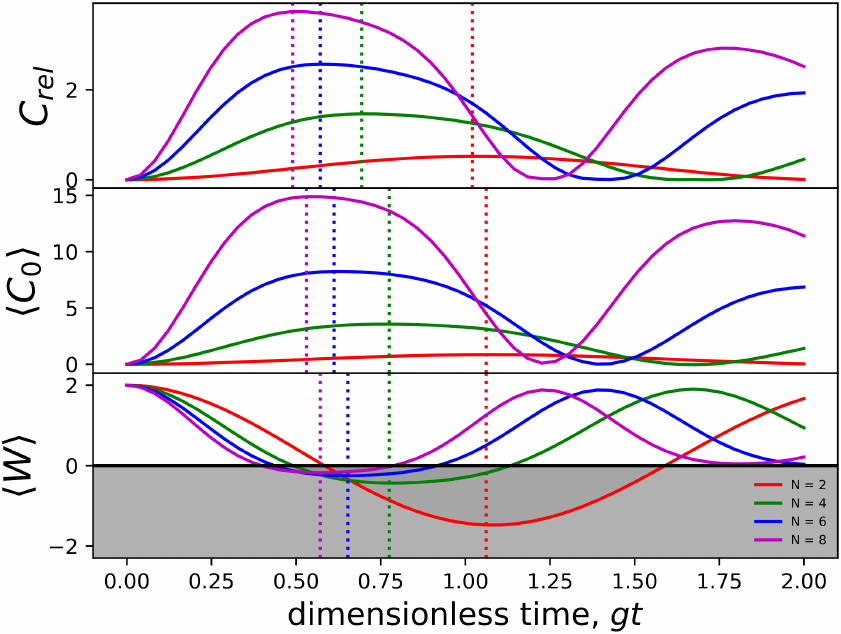}    
    \includegraphics[width=0.4\textwidth]{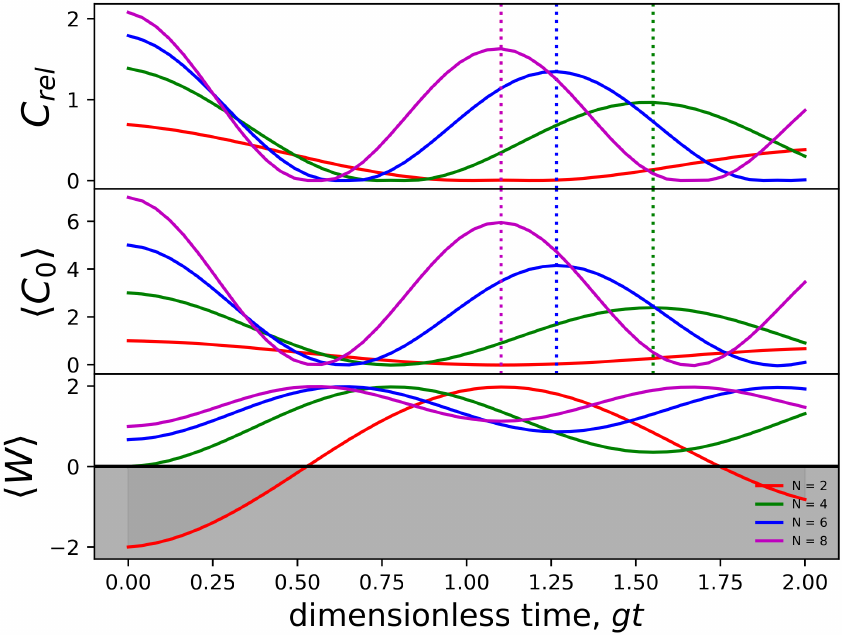} \includegraphics[width=0.4\textwidth]{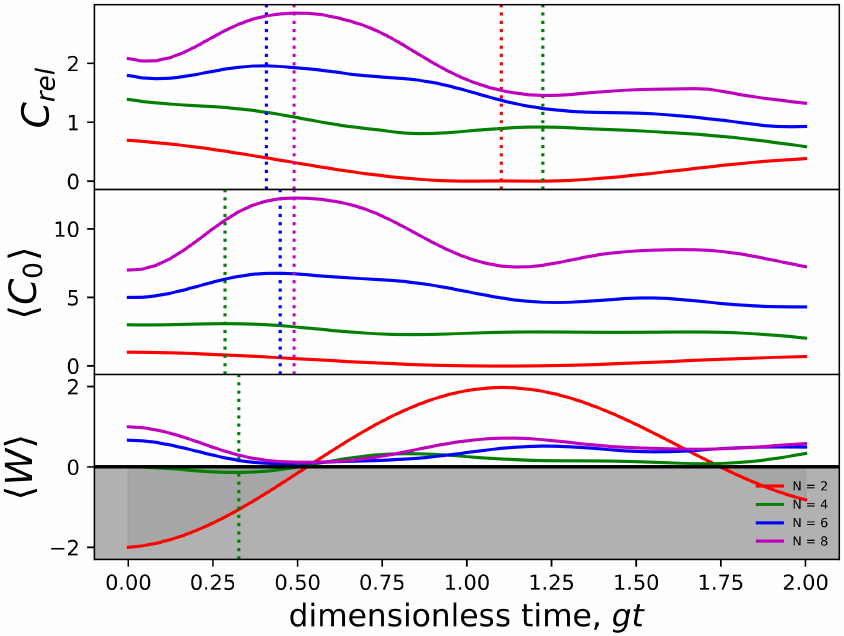} \includegraphics[width=0.4\textwidth]{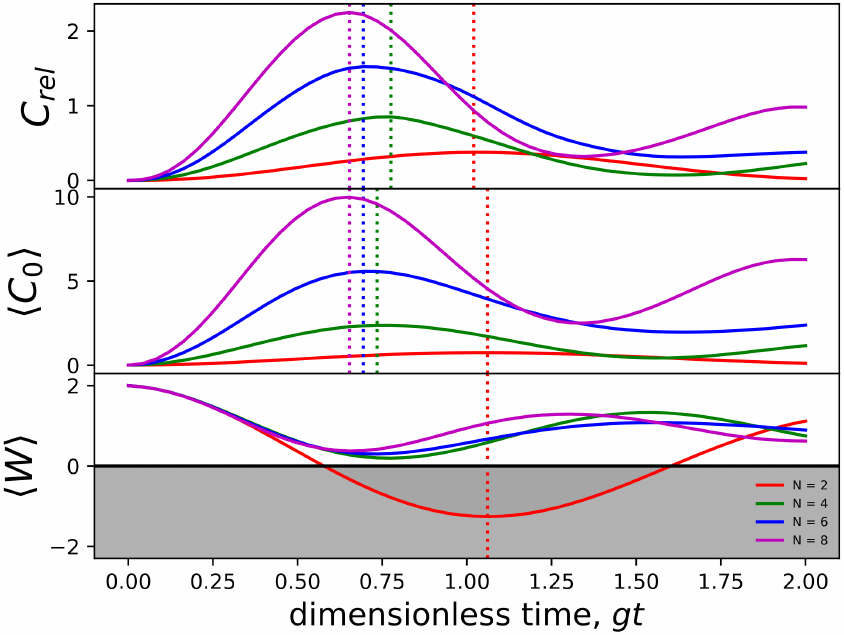} \includegraphics[width=0.4\textwidth]{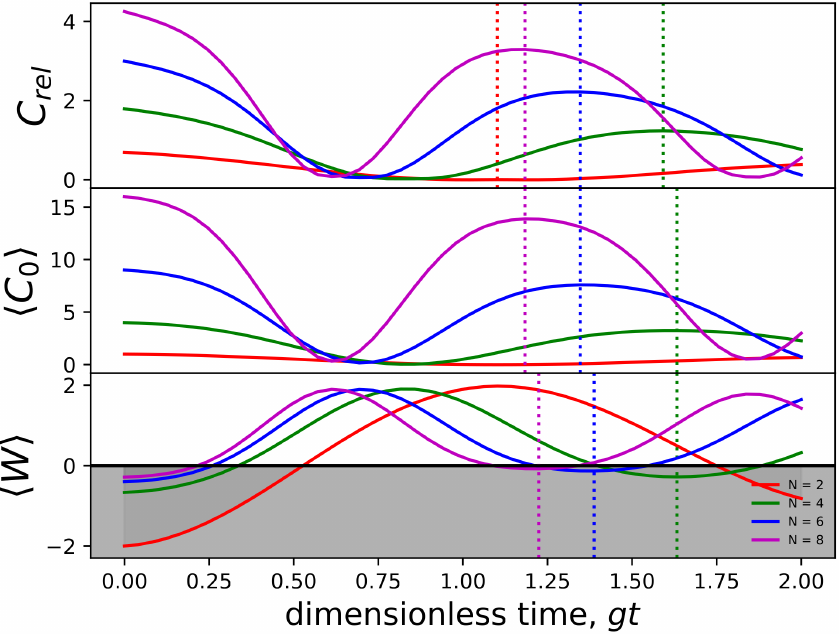} \includegraphics[width=0.4\textwidth]{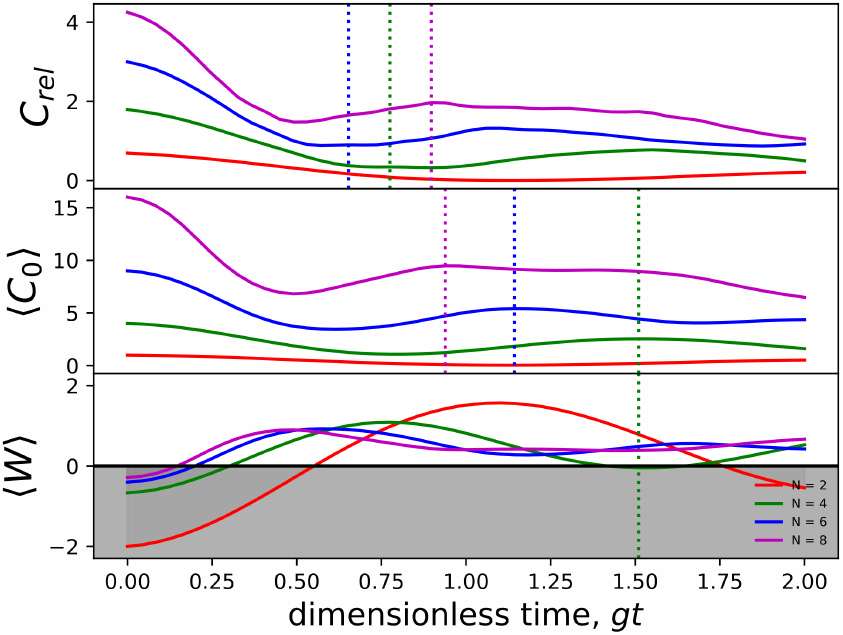} \caption{\label{fig:appendix_time_evolution} Time evolution of correlation measures for different initial states. (Top left) $N_{\text{ex}}=0$, $n_{\text{p}} = \tfrac{1}{2}\,N$. Time ordering: $\tau_{C_\text{rel}} \leq \tau_{C_0} \leq \tau_{W}$. (Top right) $N_{\text{ex}}=1$, $n_{\text{p}}=0$. Time ordering: $\tau_{C_\text{rel}}=\tau_{C_0}$. No entanglement witnessed except for $N=2$. (Middle left) $N_{\text{ex}}=N-1$, $n_{\text{p}}=0$. Time ordering: inconsistent with main article. (Middle right) $N_{\text{ex}} = N$, $n_{\text{p}} = 0$. Time ordering: ($N=2$): $\tau_{C_\text{rel}} < \tau_{C_0} = \tau_{W}$. No entanglement witnessed except for $N=2$. (Bottom left) $N_{\text{ex}} = \tfrac{1}{2}\,N$, $n_{\text{p}}=0$. Time ordering: $\tau_{C_\text{rel}} \leq \tau_{C_0} \leq \tau_{W}$. Consistent with main article. (Bottom right) $N_{\text{ex}} = \tfrac{1}{2}\,N$, $n_{\text{p}} = \tfrac{1}{2}\,N$. Time ordering: $\tau_{C_\text{rel}} < \tau_{C_0}$. Entanglement is mostly absent. } 
\end{figure*}

In Fig.~\ref{fig:appendix_time_evolution} we depict the time evolution of the coherence measures for different initial states. All other parameter remain identical:  $\gamma/g =0.1$, $\kappa/g =0.1$, $\gamma_{\phi}/g = 0.0225$, $g=1$, $\omega_{q} = 100$ and $\omega_c = 100$. In the captions $N_{\text{ex}}$ denotes the number of emitters prepared in their excited states, and $n_{\text{p}}$ is the number of photons present initially in the cavity.

\section{Effect of interaction and dissipation on separability}

Although \citet{Bassler2025} has proven that a fully inverted initial state remains separable during time evolution within the Dicke model, one may observe
entanglement in Fig.~(\ref{fig:appendix_time_evolution}). The rigorous result, however, applies only to systems with $g=0$, $\gamma=0$, $\kappa=0$ and $\gamma_{\phi}=0$.
\begin{figure*}[t] 
    \centering 
    \includegraphics[width=0.4\textwidth]{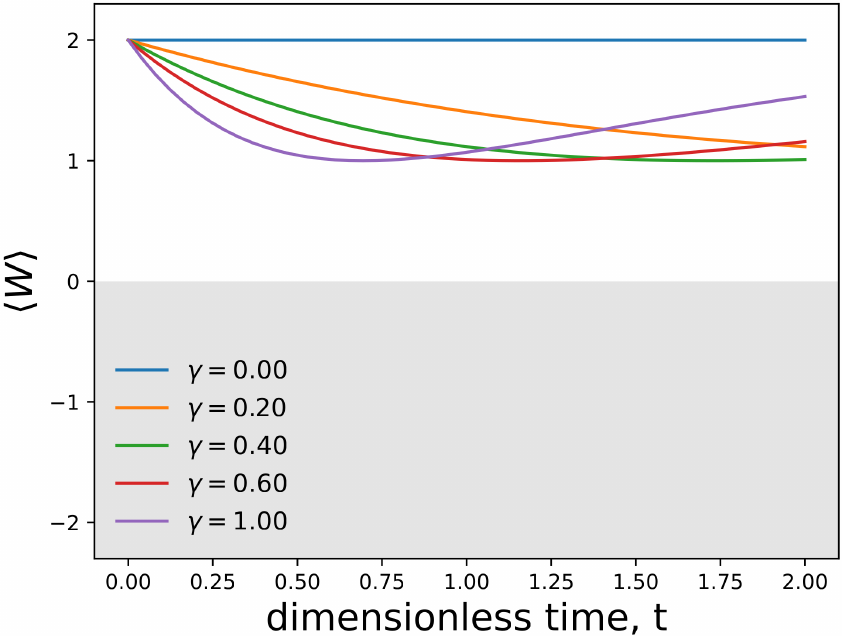} \includegraphics[width=0.4\textwidth]{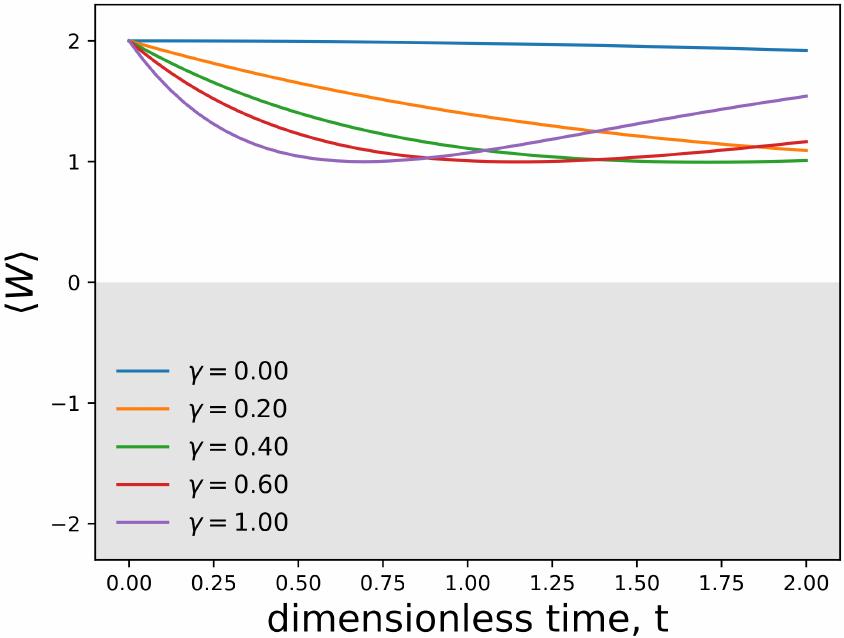} \includegraphics[width=0.4\textwidth]{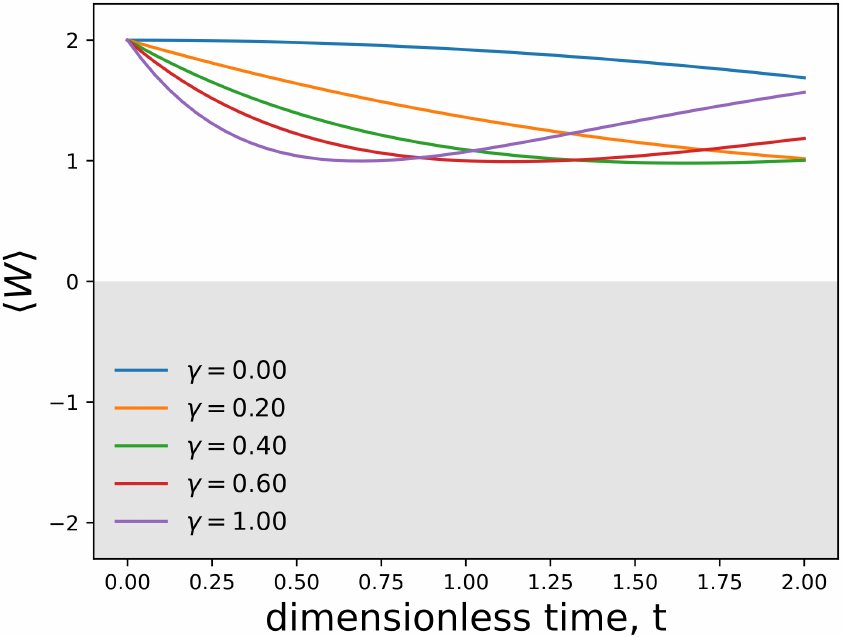} \includegraphics[width=0.4\textwidth]{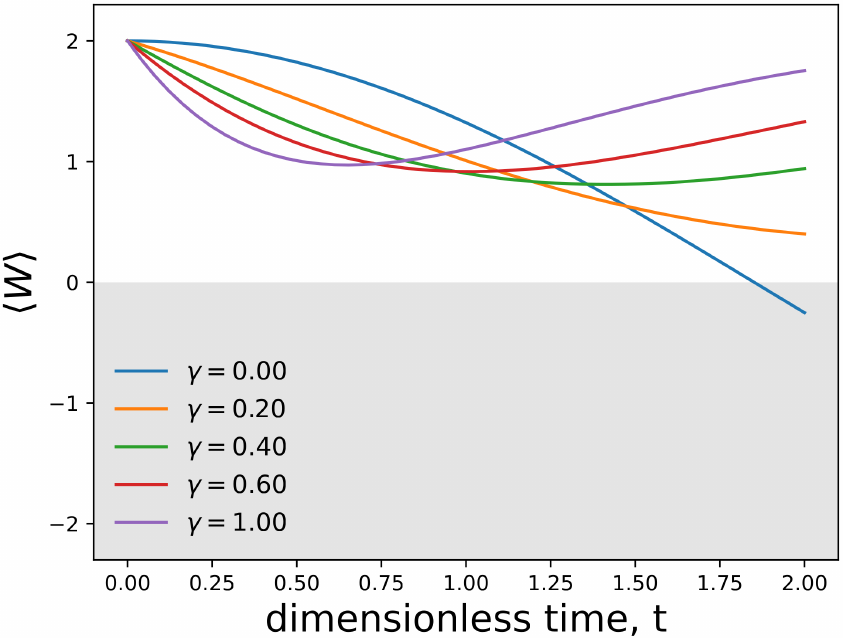} \includegraphics[width=0.4\textwidth]{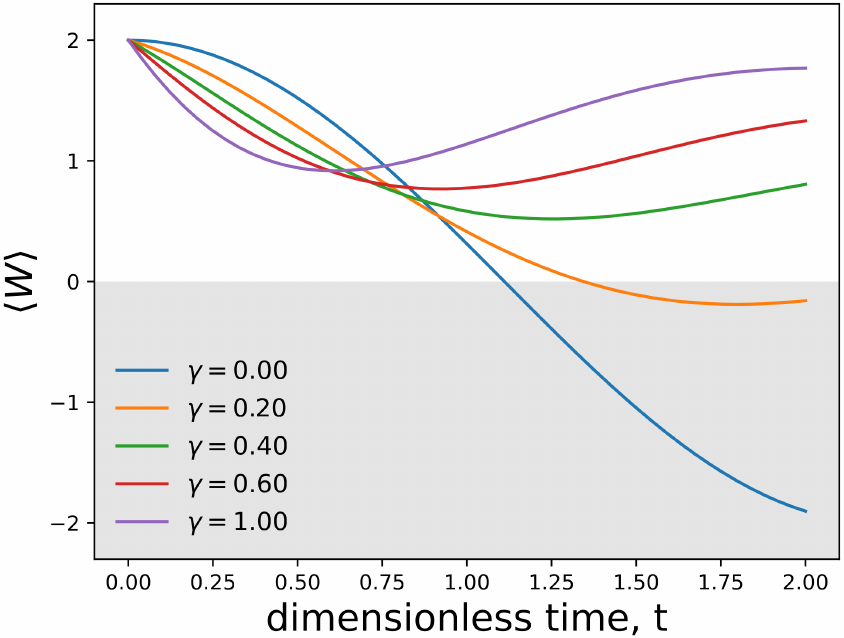} \includegraphics[width=0.4\textwidth]{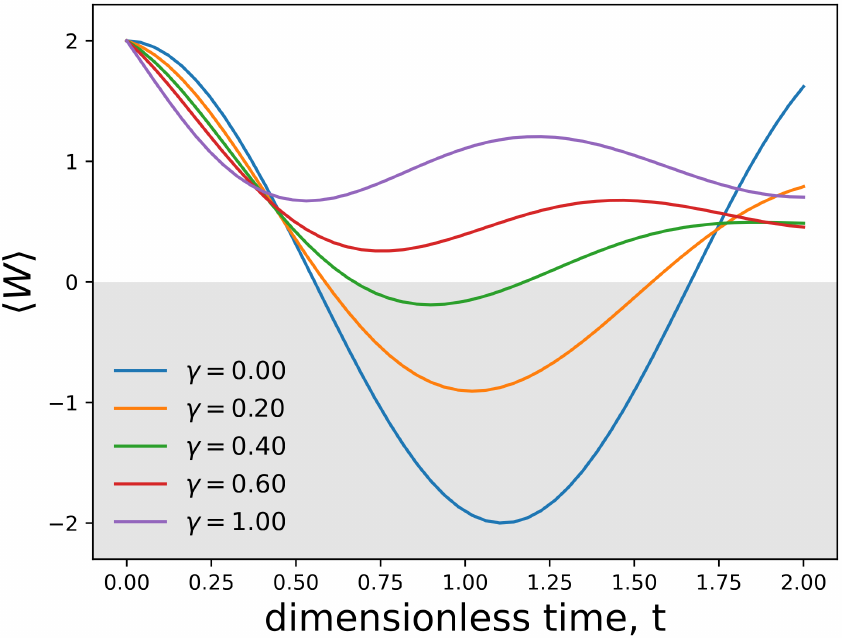} \caption{\label{fig:appendix_separability} Time evolution of entanglement witness, $\braket{W}$, for different interaction strengths, $g$, and dissipation rates, $\gamma$. (Top left) $g=0$. (top right) $g=0.05$. (middle left) $g=0.10$. (middle right) $g=0.30$. (bottom left) $g=0.50$. (bottom right) $g=1.00$. } 
\end{figure*}
Fig.~\ref{fig:appendix_separability} depicts a sequence of simulations for varying $g$ and $\gamma$, with fixed parameters $N=N_{\text{ex}}=2$, $n_{\text{p}} = 0$, $\kappa=0$, $\gamma_{\phi}=0$, $\omega_{q} = 100$ and $\omega_{\text{c}} = 100$. In each figure $g$ is fixed, while the curves correspond to different $\gamma$. One may notice lines curving downwards as $g$ increases, and even for $t<2$ and $g=0.10$, the entanglement witness, $\braket{W}$, dips below zero. However, that is not the case for $N>2$ (see Fig.~\ref{fig:appendix_time_evolution}), and we may conclude that for $g>0$ the $N=2$ case would develop entanglement.

\bibliographystyle{apsrev4-1}

%

\end{document}